\documentclass[twocolumn,prl,aps,superscriptaddress]{revtex4}
\usepackage[dvips]{graphicx}

\begin{document}

\title{Macroscopic Anisotropy and Symmetry Breaking in the Pyrochlore
Antiferromagnet Gd$_{2}$Ti$_{2}$O$_{7}$}
\author{A. K. Hassan}\email[corresponding author: ]{hassan@polycnrs-gre.fr}
\affiliation{Grenoble High Magnetic Field Laboratory, MPI-FKF and CNRS, B.P.166, 38042 Grenoble cedex 9, France}
\author{L. P. L\'{e}vy}
\affiliation{Grenoble High Magnetic Field Laboratory, MPI-FKF and CNRS, B.P.166, 38042 Grenoble cedex 9, France}
\affiliation{Institut Universitaire de France and Universit\'e J. Fourier, B.P.  41, 38402 St.  Martin d'H\`{e}res, France}
\author{C. Darie}
\affiliation{Laboratoire de Cristallographie, 25 Ave. des Martyrs, 38042 Grenoble cedex 9, France}
\author{P. Strobel}
\affiliation{Laboratoire de Cristallographie, 25 Ave. des Martyrs, 38042 Grenoble cedex 9, France}
\date{\today}

\begin{abstract}
In the Heisenberg antiferromagnet $\mathrm{Gd}_{2}\mathrm{Ti}_{2}\mathrm{O}_{7}$,
the exchange interactions are geometrically frustrated by the pyrochlore lattice structure.  This ESR study reveals a strong temperature dependent anisotropy with respect to a [111] body diagonal below a temperature $T_{A}=80$ K, despite the
\textit{spin only} nature of the $ \mathrm{Gd}^{3+}$ ion.  Anisotropy and symmetry
breaking can nevertheless appear through the superexchange interaction.  The
presence of short range planar correlation restricted to specific Kagom\'{e} planes
is sufficient to explain the two ESR modes studied in this work.
\end{abstract}

\maketitle


In antiferromagnets with competing interactions, no single spin-configuration
realizes a local energy minimum for all exchange bonds:  exchange interactions are ``frustrated''. Among all the possible sources of frustration, the most common one
has a geometrical origin. On a number of lattices (Kagom\'{e} \cite{Kano53}, fcc, pyrochlore \cite{Ramirez94}), magnetic ions are located at the vertices of equilateral triangles: the product of exchange interaction on any closed path
being negative, these structures are naturally frustrated. In the two-dimensional (2D) Kagom\'{e} lattice, equilateral triangles are connected only
through their vertices leaving a number of spin-degree of freedom unconstrained. Similarly, in the pyrochlore
lattice, spins are at the vertices of tetrahedra with equilateral triangular
faces. This lattice offers little additional constraints since all the tetrahedra
are connected in the three-dimensional (3D) structure through their corners. 
This confers a very high degree of degeneracy to the ground state (GS)
manifold: for classical spins, one or more families of continuous rotations leave
the energy of any classical ground state unchanged. As a consequence, smaller
anisotropic interactions select a particular GS through a mechanism which is
specific to the material considered. For example, a variety of ground states have been identified \cite{Ramirez94,Champion2001} in the pyrochlore family $\mathrm{R}_{2}\mathrm{Ti}_{2}\mathrm{O}_{7}$ where the rare earth ions R are
antiferromagnetically coupled. The crystal field anisotropy on the rare earth ion
$\mathrm{R}^{3+}$ plays an important role in the GS selection except for
$\mathrm{Gd}_{2}\mathrm{Ti}_{2}\mathrm{O}_{7}$, where the single-ion anisotropy of
the \textit{spin-only} $S=7/2$ ion $\mathrm{Gd}^{3+}$ ($4f^{7}$) is weak:
$\mathrm{Gd}_{2}\mathrm{Ti}_{2}\mathrm{O}_{7}$ is the only nearly-isotropic
Heisenberg pyrochlore antiferromagnet. Frustration depresses its 3D ordering
transition which has been observed in specific heat measurements
\cite{Raju99,Ramirez2001} at a temperature $ T_{c}=0.97$ K considerably smaller
than the Curie-Weiss temperature $\theta _{\mathrm{CW}}=-9.9$ K.

In this Letter, an electron spin resonance (ESR) study of the pyrochlore
antiferromagnet $\mathrm{Gd}_{2}\mathrm{Ti}_{2}\mathrm{O}_{7}$ is presented. Below a temperature $T_{A}\approx 80$ K much larger than $\theta _{\mathrm{CW}}$, \textit{two} strongly temperature dependent and anisotropic resonance lines are observed. They
identify a macroscopic anisotropy with respect to a specific $[111]\equiv{\hat{n}}$ body diagonal. These ESR modes are consistent with non-collinear spin correlations in the Kagom\'{e} planes perpendicular to this [111] axis below
$T_{A}$. This local order also agrees with the ordered state observed below $T_c$ by neutron scattering \cite{Champion2001}.
We propose an exchange-driven symmetry breaking mechanism involving the Hund term on the oxygen ion. This ion is in the center of Gd$^{3+}$ tetrahedra and is common to all the dominant superexchange paths between Gd$^{3+}$ ions. The resulting
superexchange bonds connecting the (111) Kagom\'{e} planes are weakened compared to
the exchange between the spins within these planes. Other anisotropic forces
(Dzyaloshinskii-Moriya \cite{Dzyaloshinskii57} exchange and dipolar) control the magnitude of the anisotropy and the final selection of the ground state.

\begin{figure}[tbp]
\centering
\includegraphics*[width=0.9\columnwidth]{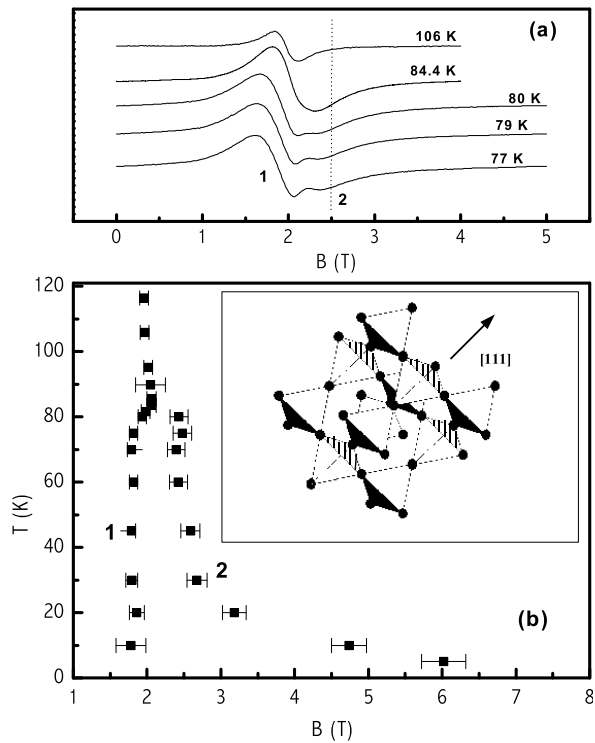}
\caption{(a) ESR spectra in the vicinity of the temperature $T_{A}$ taken at
54 GHz with the magnetic field applied parallel to the [111] axis. The
dotted line specifies the position of line 2 at 77 K after deconvolution of
the lineshape into two Lorentzians. (b) Temperature dependence of positions
of lines 1 and 2 in the same experimental conditions. At 4 K, the splitting
between the two lines (4 T $\equiv $ 5.4 K) becomes comparable to $\protect\theta _{\mathrm{CW}}$. In the inset, the corner sharing tetrahedra of the pyrochlore lattice are shown: perpendicular to the [111] axis, it can be described as a set of 2D Kagom\'{e} planes connected by interstitial spins. The filled and shaded triangles form the Kagom\'{e} planes perpendicular to the [111] axis.}
\label{fig1}
\end{figure}

The single crystals of $\mathrm{Gd}_{2}\mathrm{Ti}_{2}$\textrm{O}$_{7}$ were
grown by slow cooling from a molten flux \cite{Wanklyn79}. Crystals appeared as brown, well shaped octahedra with ca.~1~mm edge. \textrm{Gd}$_{2}$\textrm{Ti}$_{2}$\textrm{O}$_{7}$ is an insulator which crystallizes in the cubic, face centered space
group $Fd\overline{3}m$ with a lattice constant $a_{0}=10.184$ \AA\ at room
temperature. In the (111) planes, perpendicular to the cube diagonal, the Gd
spins are arranged in a 2D Kagom\'{e} lattice (see inset of figure \ref{fig1}). Only the Gd$^{3+}$ ions possess
a magnetic moment $\mu \approx 7.9\mu _{\mathrm{B}}$, close to the free-ion
value ($7.94\mu_{\mathrm{B}}$). The measured dc-magnetic susceptibility in
a field of $B_{\mathrm{ext}}=0.1$~T has a Curie-Weiss $\chi =\frac{C}{T-\theta_{\mathrm{CW}}}$ behavior typical of antiferromagnetic interactions with the same $\theta_{\mathrm{CW}}=-9.9$~K previously reported
by Raju and coworkers \cite{Raju99}. The ESR measurements were performed at
different frequencies and temperatures using a home built multi-frequency
high-field ESR spectrometer. Different frequencies were investigated (54 GHz -- 115 GHz) using back-wave oscillators and Gunn diode sources, guided to the sample with oversized waveguides in a transmission probe. The spectra were
recorded as a function of magnetic field at a fixed frequency. In this
study, we have measured (a) the temperature dependence of the ESR spectrum,
(b) its orientation dependence with respect to the crystal axes and (c) its
evolution with frequency and magnetic field.

\begin{figure}[tbp]
\centering
\includegraphics*[width=0.9\columnwidth]{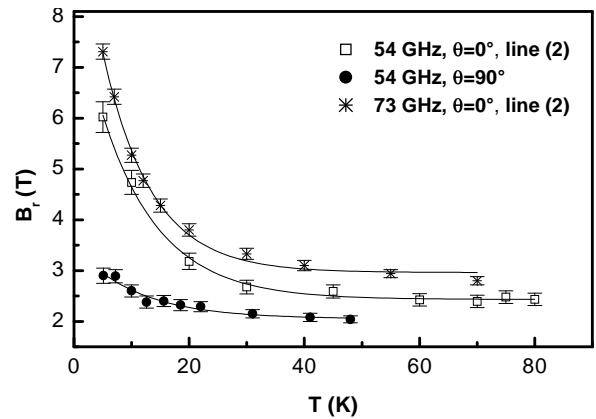}
\caption{Temperature dependence of the resonance fields. The solid lines are
fits according to the expression $B_{r}(T)=B_{0}+D\exp (-\frac{T}{T_{0}}).$
The obtained values of $T_{0}$ are 10.1$\pm $0.5 K (10.8$\pm $2.1 K) at 54~GHz for $\protect\theta =0$ ($\protect\theta =90^{\circ }$), and 8.5$\pm $0.5~K at 73~GHz for $\protect\theta =0$. At 54~GHz and $\theta =90^{\circ }$ one line is observed.}
\label{fig2}
\end{figure}

Above 80 K, the ESR spectrum at the frequency of 54~GHz consists of a single line
centered at $B_{r}\approx 2$ T, close to the expected resonance field (1.93 T) for a
$g$-factor of 2 (as expected for the \textit{spin-only} Gd$^{3+}$). As the temperature is lowered, two resonance lines appear below
$T_{A}\approx 80$~K. Their splitting is strongly anisotropic and temperature
dependent, and becomes comparable to the spin-exchange frequencies ($4 \textrm{ T} \equiv 5.4 \textrm{ K}$) at low temperature.  This is represented in figure \ref{fig1} which shows the ESR spectra at 54~GHz for different temperatures, with the magnetic field
applied parallel to a specific body diagonal of the crystal ($[111]\equiv
{\hat{n}}$ axis, perpendicular to Kagom\'{e} planes), such that the angle $\theta=\angle ({\vec{B}},{\hat{n}})=0$. Fitting the spectra below 80 K with two
Lorentzian lines (1) and (2), the position of the resonance fields can be plotted
as a function of temperature. Line (2) is strongly temperature
dependent as clearly seen in figure \ref{fig1}b. Its resonance field shifts
towards higher values as the temperature is decreased, the shift being very large ($\sim 4$~T) at low temperature (4.2~K).  As the line shifts, a broadening is also observed. While the width of line (1) also increases when lowering the temperature, the line shift is much weaker: its resonance field moves towards lower
values, the change being of the order of 0.3~T at 4.2~K. On the other hand, when the magnetic field is applied perpendicular to the body diagonal $\hat{n}$ (\textit{i.e.}\ $\theta =90^{\circ }$), at the same frequency 54~GHz, the position of the observed signal also shifts to higher fields with decreasing temperature, but the shift is smaller, $\sim 1$~T at 4.2~K. An analysis of the temperature dependence of the resonance fields below 80 K is presented in figure~\ref{fig2} for (i) both orientations $\theta =0$ and $\theta =90^{\circ }$ at 54 GHz, and (ii) for $\theta =0$ at 73 GHz.  The solid lines are fits to the phenomenological expression $B_{r}(T)=B_{0}+D\exp (-\frac{T}{T_{0}})$, (where $B_{0}$ is the resonance field at high temperature and D a parameter), consistent with an activated behavior with an activation energy $k_{B}T_{0}$.  From the fits, the obtained values of $T_{0}$ are
10.1$ \pm $0.5 K (10.8$\pm $2.1 K) at 54 GHz for $\theta =0$ ($\theta =90^{\circ }$), and 8.5$\pm $0.5 K at 73 GHz for $\theta =0$.  These values are comparable to the Curie-Weiss temperature, $\theta_{\mathrm{CW}}\approx-10$~K.

Although there is no long range order at the temperatures studied in this
work, the large temperature dependence of the position of line (2) reveals
the presence of an internal field at the frequencies probed in the
experiment. Roughly speaking, this local field adds or substracts to the
external field and produces the observed shifts. The presence of local fields
signals the growth of short range correlations as the temperature is
lowered. Since local fields fluctuations on timescales faster than the
precession period average to zero, the observed frequency shifts decrease
with increasing temperature on a scale characteristic of the short range
spin dynamics. A parallel can be drawn with ESR experiments on low
dimensional magnets where the change in resonance field as the temperature
is lowered are interpreted in terms of short range order effects \cite{Nagata72}.
\begin{figure}[tbp]
\centering
\includegraphics*[width=0.9\columnwidth]{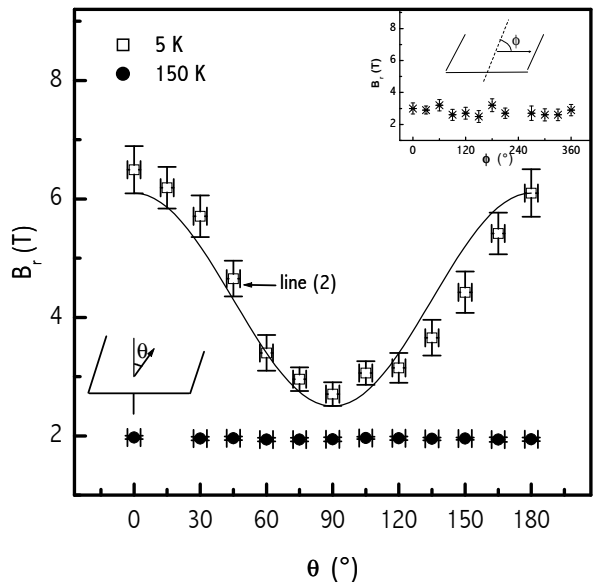}
\caption{Orientation dependence of line (2), studied at the frequency 54 GHz, as the magnetic field is rotated away from the [111] axis into the (111) plane. The strong dependence observed at 4.5 K can be fitted to $B_{r}^{0}+D\cos ^{2}\protect\theta$ with $B_{r}^{0}=2.5$~T and $D=3.6$~T. When $\protect\theta =90^{\circ}$, lines (1) and (2) merge together. The inset shows a complete rotation in the plane. At 150 K, the single resonance observed is independent of the orientation of the field.}
\label{fig3}
\end{figure}

In order to investigate further the nature of this local order, we studied the
orientation dependence of the ESR modes of our single crystal. A plot of the
resonant fields versus the angle $\theta $ between the applied magnetic field and the body diagonal $ \hat{n}$ is shown in figure \ref{fig3}, at fixed frequency (54 GHz) and for two different temperatures, 150~K and 4.5~K. The single resonance observed at
150~K is independent of the magnetic field orientation and therefore isotropic.
However at 4.5 K, the position of the resonance (2) has a strong dependence on the field direction.  As the magnetic field is rotated away from the [111] direction of the crystal, the resonance position decreases until it reaches a
minimum at $\theta =90^{\circ }$ and recovers its $\theta =0$ value as the field is
rotated towards $180^{\circ }$ with respect to the [111] direction. Hence, the
anisotropy has a uniaxial character, with an angular dependence for the resonance field 
$B_{r}^{(2)}(\theta )=B_{r}^{0}+D\cos ^{2}\theta$ (with $B_{r}^{0}=2.5$ T and $D=3.6$ T). This minimum in the resonance field $B_{r}^{0}$ occurs when the magnetic field is in the (111) Kagom\'{e} plane $\perp \hat{n}$. This singles out a unique anisotropy axis and establishes the macroscopic nature of the anisotropy. This is further evidenced by the inset of figure
\ref{fig3}, where a complete rotation $\phi$ in that plane is presented.  The dependence of the resonance fields on the orientation of the magnetic field is weaker, $\Delta B_{r,\max }$ $\sim $ 0.7 T, and may be attributed to dipolar and/or
demagnetization effects.

\begin{figure}[tbp]
\centering
\includegraphics*[width=0.9\columnwidth]{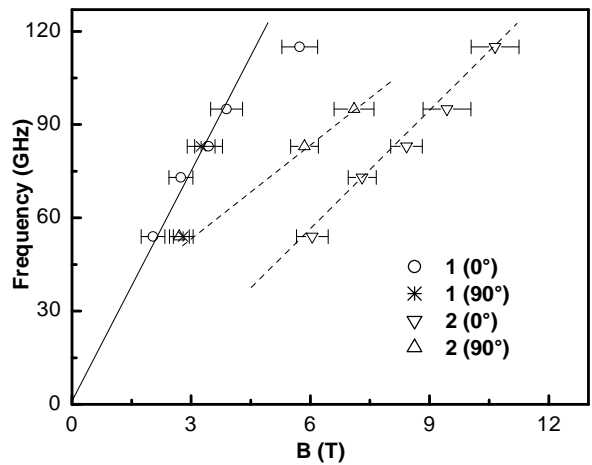}
\caption{The frequency vs field diagram for resonances (1) and (2) at 4.5 K,
parallel ($\protect\theta =0^{\circ }$) and perpendicular ($\protect\theta
=90^{\circ}$) to the body diagonal $\hat{n}$ ([111]).}
\label{fig4}
\end{figure}

The frequency dependence of resonances (1) and (2) has been studied at $T = 4.5$~K for two different orientations of the magnetic field, parallel ($\theta =0^{\circ }$, perpendicular to the Kagom\'{e} plane) and perpendicular ($ \theta =90^{\circ }$, in the Kagom\'{e} plane) to the body diagonal ($\hat{n}$). This frequency dependence
is shown as a function of magnetic field in figure \ref{fig4}. The frequency dependence
of line (1) goes through the origin and is linear with magnetic field ($\omega_{1}=\gamma H$) with a gyromagnetic ratio close to $\gamma \approx \mu /\hbar$: this is the uniform precession mode. On the other hand, line (2) has a
radically different field dependence. Considering the nature of the observed
anisotropy, the spins in the (111) Kagom\'{e} planes are not magnetically equivalent to
the spins in between them. In these circumstances, the presence of two lines in
the ESR spectrum is not really surprising \cite{Schultz80} especially if we consider the short range order present at the temperature of the measurement ($T=4.5$~K).
Hence, the resonance (2) can be assigned to a collective mode involving the
strongly correlated spins in the Kagom\'{e} plane. In the presence of
non-collinear correlations, these spins define a triad whose hydrodynamics may be
described using the angles specifying its orientation with respect to fixed lattice axes.

\begin{figure}[t]
\centering
\includegraphics*[width=0.9\columnwidth]{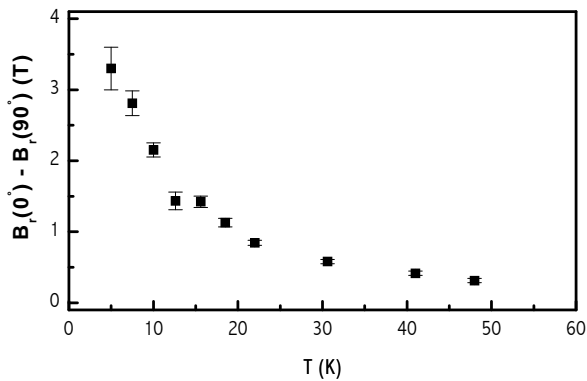}
\caption{Difference in resonance field for line (2) for a field along and
perpendicular to the anisotropy axis $\hat{n}$.}
\label{fig5}
\end{figure}

The main experimental findings may be summarized as follows:  below $T_{A}$, a
macroscopic anisotropy appears with respect to a broken symmetry axis $ \hat{n}$
lying along a specific body diagonal of the crystal. This macroscopic character
has been checked through dc-torque measurements \cite{hassan2}. This anisotropy develops
gradually as the local order (due to short range correlations) sets in when the temperature is lowered. This is
illustrated in figure \ref{fig5}, where the difference between the resonance fields
of line (2) for $\theta =0^{\circ }$ and $\theta =90^{\circ }$ is plotted versus
the temperature.  Anisotropies of local origin, such as g-factor and single-ion, are usually
unaffected by spin-correlations and thus temperature independent. Below the
ordering temperature at $T_{c}=0.97$ K, the structure of the ordered state has been
identified in a neutron scattering experiment \cite{Champion2001} with an isotope
enriched $^{160}\mathrm{Gd}_{2}\mathrm{Ti}_{2}\mathrm{O}_{7}$ sample: within the
Kagom\'{e} plane, spins are ordered in a chiral $120^{\circ }$
structure (the same as the ``$q=0$'' spin structure observed in ordered Kagom\'{e} antiferromagnets \cite{Reimers93}), while the structure for the spins in between the
planes has not been definitively established. This structure is fully consistent with the anisotropic local order observed here at higher temperatures. M\"{o}ssbauer experiments also conclude in a planar structure for the
spins in the Kagom\'{e} planes below $T_{c}$ \cite{Bonville2001}.

We now consider the microscopic forces which could give rise to this
macroscopic anisotropy with respect to a specific body diagonal. Let us
first consider the main super-exchange interaction between neighboring Gd$^{3+}$ ions. Two oxygen ions bridge the vertices of the tetrahedral Gd$^{3+}$ pyramid. The shortest bridge goes through the oxygen located in the center
of the tetrahedron, which bridges \textit{all} four Gd$^{3+}$ ions at the
tetrahedron vertices.  We focus on this shortest bridge. If
only the strong ligand electric fields are considered, the four oxygen
orbitals are identical $sp^{3}$ mixtures pointing toward the tetrahedron
vertices. In a $T_{d}$ symmetry, the Hund term on the oxygen splits the four
$sp^{3}$ orbitals into a singlet and a triplet \cite{Levy}. This energy splitting will
reduce the superexchange coupling between the Gd$^{3+}$ ion coupled to this
singlet $sp^{3}$ oxygen orbital with respect to all other three Gd$^{3+}$ ions.
If all singlet $sp^{3}$ oxygen orbitals belonging to different tetrahedra
``lock'' along the same $[111]$ body diagonal, the system has chosen the
appropriate broken symmetry axis. This form of ``\textit{orbital ordering}'' would
naturally occur at a much higher temperature ($T_{A}\approx 80$~K) than the
typical exchange coupling measured by the Curie-Weiss temperature, $\theta _{\mathrm{CW}}\approx -10$~K. This
mechanism by itself does not provide a spin-anisotropy. However, the
Dzyaloshinskii-Moryia exchange interactions \cite{Dzyaloshinskii57} will
provide in this circumstance a macroscopic coupling $D_\mathrm{DM}^{ij}{\hat{n}}\cdot {\vec{S}}_{i}\times {\vec{S}}_{j}$ for the spins ${\vec{S}}_{i}$ and ${\vec{S}}_{j}$ in the Kagom\'{e} plane. For a pure Kagom\'{e} lattice, this
term is known \cite{ElHajal2002} to give rise to the same chiral
order as observed by neutron scattering \cite{Champion2001}. Dipolar
interactions \cite{Palmer2000} could also favor such a state, once the spins
are restricted to the Kagom\'{e} planes. In order to experimentally validate
this scenario, the knowledge of the magnetic structure for the different
magnetic phases in a magnetic field \cite{Ramirez2001} would be decisive.

Whether this exchange mechanism plays a significant role in other pyrochlore
magnets with large single-ion anisotropies is also open. For example,
in the $\mathrm{Tb}_{2}\mathrm{Ti}_{2}\mathrm{O}_{7}$ pyrochlore, the
crystal field splitting of the Tb$^{3+}$ ion is estimated to be in the
$15-20$ K range \cite{Gingras2000} which is far less than the temperature ($
50$ K) where short range order sets in \cite{Gardner99}. In such
circumstances, an exchange-driven anisotropy could be important.

In conclusion, this ESR study has revealed an unexpectedly large magnetic
anisotropy for a pyrochlore magnet with spin-only Gd$^{3+}$ ions. Its possible
exchange driven nature stresses the importance of the subtleties of the microscopic coupling
between the rare earth ions in these frustrated magnets where a large degree of
degeneracy is present.

We gratefully acknowledge very helpful discussions with B. Canals, M. Elhajal, C. Lacroix and A. Wills.


\begin{references}
\bibitem{hassan2} The results of the dc torque measurements will be published elsewhere.
\end{references}
\end{document}